\documentclass[12pt,preprint]{aastex}

\begin{document}

\title{The Birth of Quark Stars: Photon-driven Supernovae?}

\author{Anbo Chen, Tianhong Yu and Renxin Xu}
\affil{School of Physics, Peking University, Beijing 100871, China
}%
\email{r.x.xu@pku.edu.cn}

\begin{abstract}
In this letter we propose a possible mechanism trying to alleviate
the current difficulty in core-collapse supernovae by forming a
strange quark star inside the collapsing core.
Although the initial longtime cooling behavior of nascent strange
stars is dominated by neutrino emissions, thermal emissions
including photons and $e^\pm$ pair plasma do play a significant
role in the explosion dynamics under this picture.
The key to promote a successful shock outside a bare strange star
is more likely to be the radiation pressure caused by thermal
photons rather than neutrinos in conventional models.
We observed through calculation that radiation pressure can push
the overlying mantle away through photon-electron scattering with
energy (the work done by radiation pressure) as much as $\sim
10^{51}$ erg if protoquark stars are born with temperatures higher
than $\sim (30-40)$ MeV.
This result not only indicates that strange quark stars should be
bare ever since their formations, it could also provide a possible
explanation to the formation of fireballs in cosmic long-soft
$\gamma$-ray bursts associated to supernovae.
\end{abstract}


\section{Introduction}

Thanks to the advanced X-ray missions (e.g., {\em Chandra} and
{\em XMM-Newton}), it is now high time for astrophysicists to
research into the nature of pulsars and relevant issues.
Pulsar-like stars are unique astro-laboratories to study matter at
supranuclear density.
On the one hand, due to the mathematical complexity of the
nonlinear nature of quantum chromo-dynamics (QCD), one believed to
be the underlying theory for strong interaction, we can not
determine the state of supranuclear matter by first principles.
Several speculations have to be presented in the literatures,
including those currently focused state-of-the-art nuclear
equations (normal neutron stars). Besides these conjectures,
stable strange quark matter state (quark stars) has been
alternatively proposed since quark matter is a direct consequence
of the ``asymptotic freedom'' which was found experimentally in
1960s and proved by QCD in 1973.
On the other hand, recent observations show new members of the
family of pulsar-like stars (e.g., anomalous X-ray pulsars/soft
gamma-ray repeaters, central compact objects, and dim thermal
neutron stars), whose different manifestations are not well
understood and could challenge the conventional scenario of normal
neutron stars.
In fact, neutron stars and quark stars should now be considered as
two potential models equally possible for the nature of pulsar-like
stars \citep[see reviews
e.g.,][]{madsenreview,lp04,glen,weber05,xu06}.

How do quark stars form?
This is a question with bimodality of meaning.
(i) It is straightforward to know their births if one believes
pulsars are actually quark stars.
The astrophysics of phase conversion from nuclear matter to quark
matter during, e.g., spindown or accretion stages is investigated at
a preliminary step \citep{odd02,koj05}; but that quark star
formation occurs simultaneously during the collapsing of massive
star's cores has not been treated previously, which will be focused
in this paper.
(ii) Could core-collapse-produced quark stars result in successful
supernovae?
While supernovae keep occurring above the sky, the failure to
simulate an explosion successfully in calculations troubles
astrophysicists over time. In reviewing the neutrino-driven
explosion model, the call for an alternative mechanism grew
stronger~\citep{Mezzacappa05,Buras:2003PRL}.
Since strange quark matter (SQM) could be the real ground state
\citep{Witten:1984PRD}, it is suggested that SQM-formation may
help to overcome the energy difficulty in getting type-II
supernovae successful, because more neutrinos should be radiated
if phase-transition to SQM is included
\citep{bh89,Lugones:1994PRD,dpl95,Horvath2007}.

However, it should be emphasized that bare quark surfaces could be
essential to successful explosions, for both $\gamma$-ray bursts
and core-collapse supernovae \citep{ph05,xu05}.
The reason for that is simple and intuitive: due to the chromatic
confinement, the photon luminosity of a quark surface is not
limited by the Eddington limit.
Regarding the ultra-high surface temperature of nascent strange
stars~\citep{Haensel:1991APJ}, we believe that the strong radiation
pressure caused by enormous thermal emissions from strange stars
might plays a more important role in promoting a shock, substituting
neutrinos in conventional delayed shock mechanism.
We call this proposed scenario as a {\em photon-driven supernova}.
Quark stars formed in this way should be bare, and factually,
there are possible observational evidence for bare strange stars
\citep{xzq99,xu02,yue06}.

\section{The Model}

Due to the different properties between neutron stars and strange
stars, the consequent mechanism in explosion could be significantly
altered.
A detonation wave burning nuclear matter into strange matter spreads
out from the core~\citep{Lugones:1994PRD}, where the density
declines as radius increases. A boundary of strange matter and
nuclear matter will be set where the detonation wave stops when
nuclear matter density drops blow a critical value. Nevertheless,
the outer mantle on the other hand is still in-falling with hardly
any variation since the detonation wave travels faster than the
sound speed. In both the prompt and the delayed shock mechanisms,
the previous shock which ejects the overlying mantle is initially
generated by the bounce at the surface of neutron
stars~\citep{Bethe:1990RMP}. However, if a nascent strange star is
formed by a detonation wave, the strange matter behind the wavefront
is considered nearly at rest~\citep{Lugones:1994PRD} and hence no
bounce could be introduced.

As a consequence of the high optical opacity in the collapsing
core with density $\rho$ and temperature $T$ \citep{Pad:book},
$%
\kappa_{\rm ph}=\kappa_{\rm ph_0}{\rho}T^{3.5}~{\rm cm^2g}^{-1}
$, %
the radiation pressure caused by photon-electron scattering
prevents a thin layer of nuclear matters outside from falling onto
the surface of nascent strange star and thus inverses its
velocity, where $\kappa_{\rm ph_0}=4\times10^{25}Z(1+X)$, atomic
charge number $Z$, and hydrogen mass fraction $X$. This layer
being affected extends to the sonic point because super-sonic
fluid cannot be affected by perturbations downstream.
As soon as the layer between the strange star surface and the
sonic point assembles the radiation pressure after an equilibrium
is established, a shock will be generated at the sonic point for
the discontinuities in both density and pressure.

\begin{figure}
\plotone{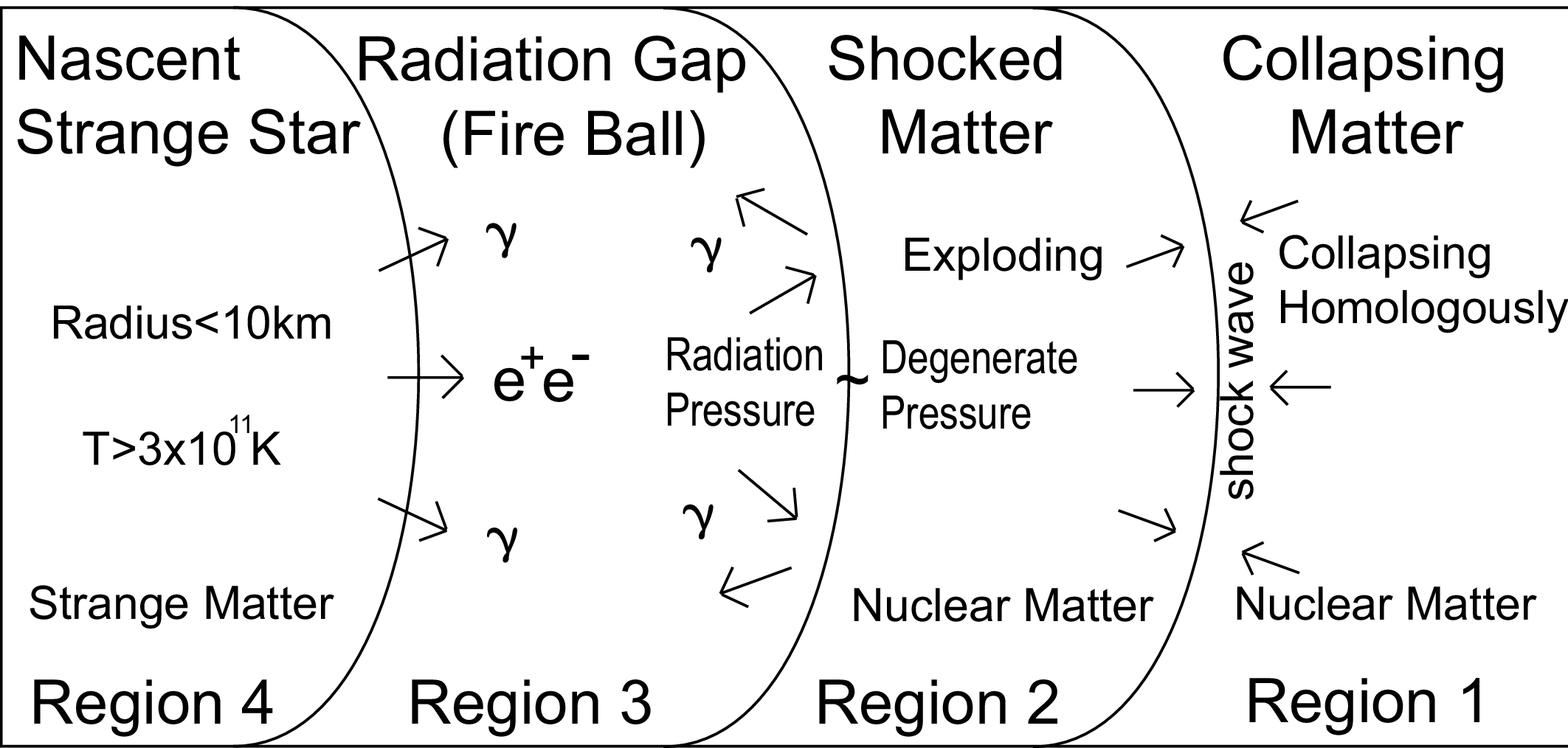} \caption{\label{fig1:Illustration}The outermost
region 1 consists of the unshocked nuclear matter which is still
in-falling, assembled to the homologous
solution~\citep{Goldreich:1980APJ,Yahil:1983APJ}. Behind the shock
front which serves as the border and increases in thickness,
region 2 comprises the shocked nuclear matter whose motion has
been reversed by the shock. Between the nascent strange star in
the center of the original collapsing core and region 2 is a
fireball (region 3), a gap filled up with high energy photons and
$e^+e^-$ pair plasma, similar to the fireball in case of long-soft
gamma-ray bursts.}
\end{figure}

As the overlying mantle is driven outwards by the shock, a gap is
left between the strange star and the out-going matter.
This gap is filled up by a {\em fireball}, with high energy
photons and electron-positron pair plasma emitted from the hot
strange star. We illustrate the initial collapsing core divided
into four parts in Fig.~\ref{fig1:Illustration}.

Though annihilation of $\gamma$ into $e^+e^-$ pairs and the reverse
process take place at such high energy scale, the total momentum and
energy fluxes are preserved. Hence we may still adopt the formula
$P_{\rm rad}=aT^4/3$, where $P_{\rm rad}$ is radiation pressure and
$a$ is the radiation constant.
The pressure decreases as the fireball expands, with an assumption
of $P_{\rm rad}\propto r^{-n}$, where $r$ is the radius of
fireball. In a radiation dominated relativistic fireball with a
vacuum exterior, we have $n=4$, i.e., $P_{\rm rad}\propto r^{-4}$
~\citep{Piran:1993}. However, the exterior of the fireball is
actually not vacuum but fall-back matter, which must exert a force
to the fireball. The expanding speed of fireball is lower than the
speed of light, and the radiation pressure should then not drop so
fast, i.e. the index $n$ could be smaller than 4. In a special
case for normal radiation, one may have $P_{\rm rad}\propto
r^{-2}$ from flux conservation. Therefore we generalize the
relation as
\begin{equation} P_{\rm
rad}(r)=\frac{1}{3}aT^4\frac{R^n}{r^n}\label{equation:Prad},
\end{equation}
where $R$ is the radius of the strange star, and three cases of
$n=4,3,2$ will be considered.

Our main goal is to check whether the work done on the overlying
mantle by the radiation pressure is comparable to $10^{51}$ erg,
the typical energy needed for a successful supernova explosion.
The work done by the fireball reads,
\begin{equation}
W=\int_R^{r_{\rm f}}4{\pi}{r^2}P_{\rm rad}(r){\rm d}r .
\end{equation}
where $r_{\rm f}$ is the final distance that the radiation pressure
could push the out layers to. The distance can be estimated by
velocity of the shocked matter and the radiation-dominated
timescale.

Since mean free path of photons in dense nuclear matter is extremely
short,
$%
\lambda=1/(\rho\kappa_{\rm ph})=1/(\kappa_{\rm ph_0}\rho^2T^{3.5})
$, %
radiative heating is very inefficient. The time scale for
transferring most part of the energy to the shock is excessively
shorter than the kinematic timescale, as we can see later in our
calculation.
The lack of energy transferred to the outside of fireball could
essentially help a successful explosion.
Moreover, we can therefore omit the thin layer that has been
heated in region 2 and consider the pressure within these two
regions contributed by degenerate electrons alone, complying with
the polytropic equations:
$%
P_1 \propto\rho_1^\gamma, P_2 \propto\rho_2^\gamma
$. %
Because both the shock and the shocked matter are supported by the
radiation pressure ultimately, we assume that pressure in region 2
equals to the radiation pressure for simplicity though in the
actual explosion non-uniform pressure distribution must exists.

Meanwhile, we solve the relativistic shock equation
\citep{Landau:book} and obtain the speed of the shocked matters
moving outwards,
\begin{equation}
\frac{v_1}{c}=[\frac{(P_2-P_1)(e_2+P_1)}{(e_2-e_1)(e_1+P_2)}]^{1/2},\\
\frac{v_2}{c}=[\frac{(P_2-P_1)(e_1+P_2)}{(e_2-e_1)(e_2+P_1)}]^{1/2}.
\label{equation:Shock}
\end{equation}
where $v$ is velocity and $e$ stands for energy density. Note that
these values are derived in a coordinate system where the shock
wave surface is at rest. Subscript 1 and 2 represent corresponding
values in regions 1 and 2 (defined in
Fig.~\ref{fig1:Illustration}), respectively. Adopting the
numerical solution of last good homology [Table 1 of
\cite{Brown:1982NPA}] and applying the radiation pressure of
Eq.(\ref{equation:Prad}) to region 2, the velocity of the shock's
propagation rate can be achieved by
\begin{equation}
V_{\rm shock}=c^2\frac{v_1-V_1}{v_1V_1-c^2},
\end{equation}
where $V$ is the velocity in the inertial frame where the
collapsing center is at rest. Thus the velocity of region 2 in the
center inertial frame can be obtained,
\begin{equation}
V_2=c^2\frac{V_{\rm shock}+v_2}{V_{\rm shock}v_2+c^2}=\frac{{\rm
d}r}{{\rm d}t}.
\label{equation:Expand}
\end{equation}
Finally, we obtain the velocity of the shocked matter (about 0.1
speed of light) and hence the expansion speed of the fireball.

Strange quark matter at high density and small temperature is
expected to exhibit color-superconductivity (CSC), induced by
quark pairing and condensation at the Fermi surface, with energy
gaps $\Delta\simeq$100 MeV \citep{Alford:1998,Ouyed2005} and
associated critical temperatures $T_{\rm c}\simeq0.6\Delta$, above
which thermal fluctuations destroy the condensate
\citep{Rajagopal:2000}. If the density is sufficiently high, a
color-flavor locked (CFL) phase is the favored ground state
\cite[e.g.,][]{Ouyed2005}.

The surface emissivity of photons with energies below
$\hbar\omega_{\rm p}\simeq$ 23MeV ($\omega_{\rm p}$:
electromagnetic plasma frequency) is strongly suppressed
\citep{usov01}. As shown in \cite{VRO}, average photon energies in
CFL matter at temperature $T$ are $\sim 3T$. Therefore, when the
surface temperature of the star drops below $T_{\rm
f}=\hbar\omega_{\rm p}/3\simeq$ 7.7MeV, the photon emissivity can
be considered shut off. As long as we know the cooling behavior,
we can then estimate the timescale of radiation-dominated period.

A nascent quark star is actually not an isothermal, homogeneous
sphere \cite[e.g.,][]{Ouyed2005}.
The thermal evolution is determined by the energy conservation and
heat transport equations,
\begin{equation}
C_{\rm v}\frac{\partial T}{\partial
t}=-\frac{1}{r^2}\frac{\partial(r^2F_r)}{\partial r}-\epsilon_\nu,\;
F_r=-\kappa\frac{\partial T}{\partial r},
\label{equation:Cool}
\end{equation}
where $C_{\rm v}$ is the specific heat of the star matter, $\kappa$
is its thermal conductivity, $\epsilon$ is the neutrino emissivity,
and $F_r$ is the heat flux at radius $r$.

For normal quark matter \citep{Iwamoto82}, one has
\begin{equation}
C_{\rm v}\simeq 1.6\times10^{21}T_9{\rm\,erg\,cm^{-3}\,K^{-1}},\;
\epsilon_\nu=3\times10^{24}T_9^6{\rm\, erg\, cm^{-3}\, s^{-1}}.
\end{equation}
The heat capacity of the quark matter with color superconductivity
\cite[CFL phase,][]{Blaschke:2000} could be
\begin{equation}
C_{\rm v}\simeq 5.1\times10^{21}T_9(T_c/T){\rm exp} (-\Delta/T)
[2.5-1.7T/T_c+3.6(T/T_c)^2] {\rm\, erg\, cm^{-3}\, K^{-1}},
\end{equation}
where $\Delta$ is the energy gap and $T_c$ is the critical
temperature, and the neutrino emissivity for this CFL state is
\citep{Jaikumar:2001},
\begin{equation}
\epsilon_\nu=2\times10^{20}T_9^7{\rm\, erg\,cm^{-3}\,s^{-1}}.
\end{equation}
It is evident from Eq.(7) and Eq.(9) that the neutrino emissivity
could be much lower in CFL state than in normal state of quark
matter. Low $\nu$-emissivity favors keeping a high surface
temperature of CFL stars (Fig. 2), that helps to form an energetic
fireball. According to \cite{Shovkovy:2002}, the conductivity in
Eq.(\ref{equation:Cool}) reads
\begin{equation}
\kappa=1.2\times10^{27}T_{\rm MeV}^3\lambda_{\rm GB}~{\rm\, erg\,
cm^{-1}\, s^{-1}\, K^{-1}},~~~~~\lambda_{\rm
GB}(T)=\frac{4(21-8{\rm ln}2)}{15\sqrt 2 \pi T_{\rm MeV}}~{\rm
cm}.
\end{equation}

In addition, we choose the following boundary conditions: at the
center and the surface of the star, we have $F(r=0)=0$ and
$F(r=R)=\sigma T^4$. As for the initial condition, we assume the
temperature of the star at $t=0$ be uniform, $T(t=0)=T_0$. The
initial temperature of the new born strange star can be estimated
to be tens of MeV. Once we input $T_0$, Eq.(\ref{equation:Cool})
can be solved numerically. We can then obtain radiation-dominated
timescale and hence calculate the total work done by the fireball.

\section{The Results}

\begin{figure}
\plotone{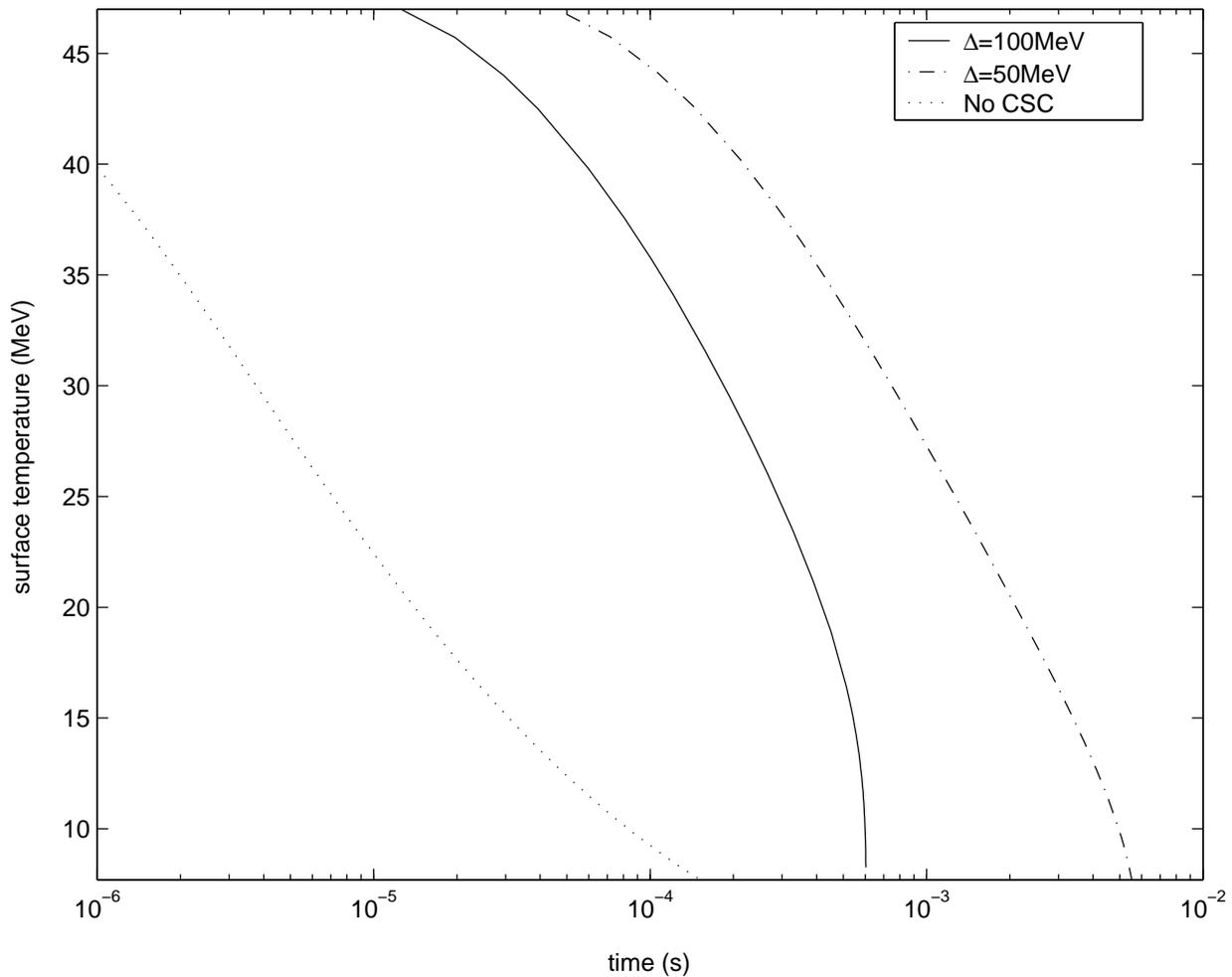} \caption{\label{fig2:Cooling}Cooling of a strange
star with radius $R$=9 km and initial temperature $T_0$=50 MeV,
for two different energy gaps of CFL matter, $\Delta$=100MeV
(solid line), 50MeV (solid-dotted line), and for normal quark
matter (dotted line). The cooling of CFL matter is slower than
normal quark matter, and the CFL matter cools faster with bigger
energy gap. In about $\sim$1 ms, the surface temperature of a CFL
star will drop below 7.7 MeV, below which photon emission would
not be effective.}
\end{figure}

In the calculation shown in Fig.~\ref{fig2:Cooling}, we find that
the cooling rate of strange matter with CSC is slower than that of
normal quark matter, and with bigger energy gap, the CSC matter
cools faster. In about $\sim$1 ms, the surface temperature of a
CFL star will drop below 7.7 MeV, the typical threshold of photon
emission. The timescale of strong thermal radiation is adequately
short, and thus we can safely use the assumption that the heating
process in the thin layer in region 2 of
Fig.~\ref{fig1:Illustration} will not influence much to the
mechanism.

\begin{figure}
\plotone{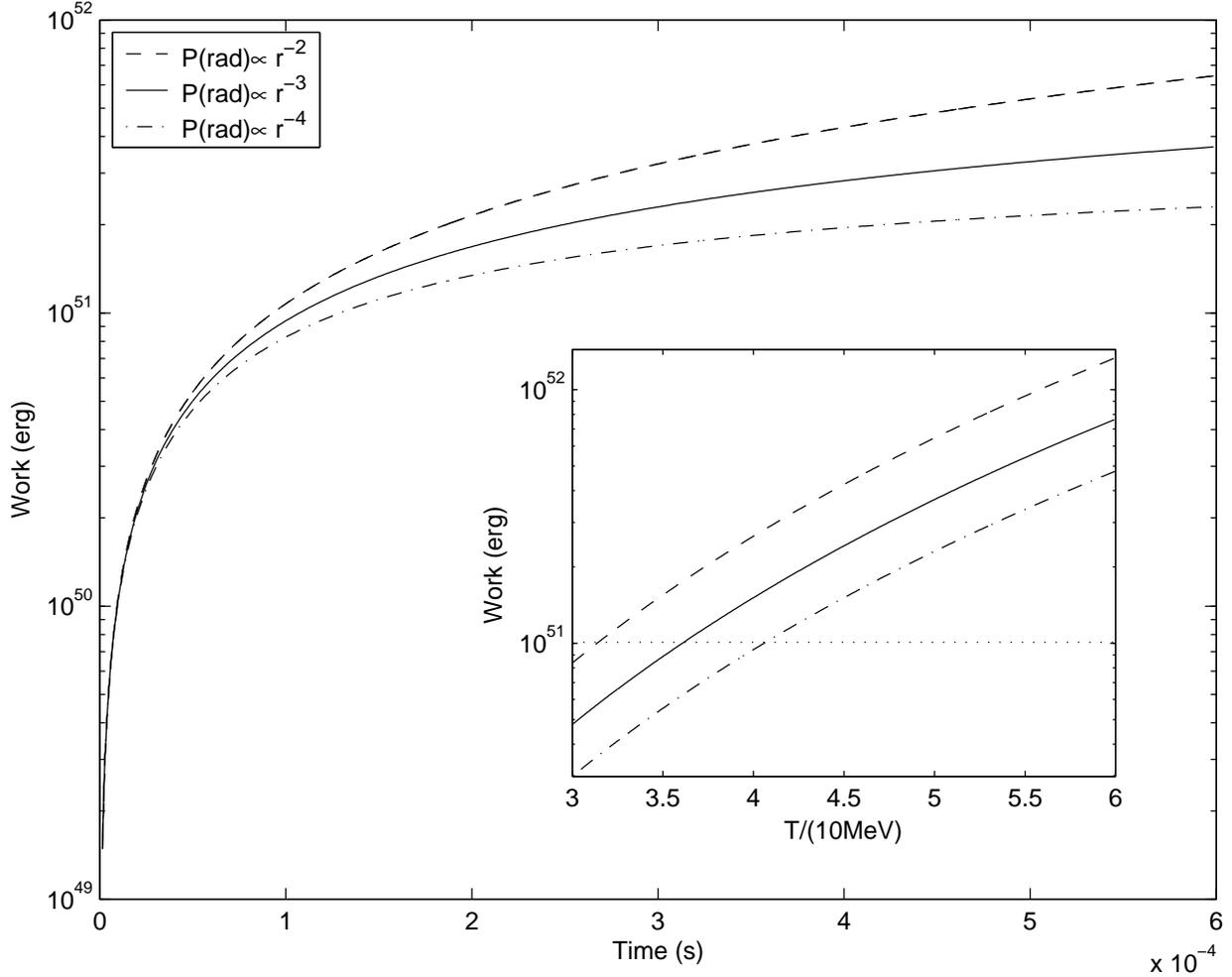} \caption{\label{fig3:Work}Work done by photon on
the overlying mantle, with different radiation pressure-fireball
radius relationship and different initial temperatures. Both figures
are under Strange Star model with radius R=10km. Different lines
reflect different values of index n in Eq.(1). The larger figure
shows the work increasing with time, which is calculated using
initial temperature $T_0=50 {\rm MeV}$. The smaller figure at
right-down corner shows the work with different initial
temperatures, within radiation dominated timescale $t\sim 0.0006 s$.
The work could exceed $10^{51}$ erg if initial temperature is higher
than about 40 MeV. }
\end{figure}

In Fig.~\ref{fig3:Work} we calculate the work done by radiation
pressure with different pressure-radius relationships, i.e.
different value of index $n$ in Eq.(1), and with different initial
temperatures. It can be seen that, for a typical condition (e.g.,
$T_0=50 {\rm MeV}, R=10 {\rm km}, n=3$), the work would exceed
$2.7\times 10^{51}$ erg within radiation dominated timescale,
which may result in a successful explosion. For different initial
temperatures, the work could pass $10^{51}$ erg once the initial
temperature exceed about 40MeV.
Note that the initial temperature could be (30-40) MeV for normal
neutron stars \citep{BL86}. More energy (i.e., both gravitation
and phase-transition energies) should be released for quark stars
if the conjecture by \cite{Witten:1984PRD} is correct.
Unfortunately, it is still model-dependent to determine the real
initial temperature of a quark star. From Fig.3, we could conclude
that the total work done by the fireball could be enough for
successful supernova if the initial temperatures is greater than
$\sim (30-40)$ MeV.

\section{Conclusions and Discussions}

A photon-driven mechanism is proposed for successful core-collapse
supernovae, and bare strange quark stars are residues after the
explosions.
It is found through calculations that radiation pressure can push
the overlying mantle away through photon-electron scattering with
energy (the work done by radiation pressure) as much as $\sim
10^{51}$ erg, the typical energy needed for core-collapse
supernovae.

Regarding the difference between nuclear matter and strange quark
matter, a strange star can have a much higher surface temperature
than a neutron star and accordingly a greater thermal photon
emission.
In addition, the strong electric field ($\sim 10^{17}$ V/cm) on
quark surface should play an important role in producing the
thermal emission too [i.e., the Usov mechanism
\citep{usov98,usov01}], which is in the same order of blackbody
radiation when $T>5\times 10^{10}$ K.
Furthermore, this distinction results in a huge radiation pressure
that leads to a much faster explosion than the conventional
delayed-shock model. The photon-driven supernova may benefit not
only from the radiation pressure but also a much smaller
photo-dissociation effect while most part of the mantle is blown
away before the iron-cores could ever interact with high energy
photons, hence making it possible to provide sufficient energy and
promote a successful explosion.

Low-mass bare strange stars could form via accretion-induce
collapse of white dwarfs \citep{xu05}. A low energy budget is
needed in this scenario since the gravitational binding energy of
a white dwarf with approximate Chandrasekhar mass is only $\sim
10^{50}$ erg.

This photon-driven mechanism may also provide an alternative
figure how a fireball can be produced during cosmic long-soft
$\gamma-$ray bursts, which are observed associated with
supernovae.
It is worth noting that, due to the chromatic confinement of quark
surfaces, the baryon contamination would be very low in such
fireballs, that is necessary in the models
\citep{cd96,dl98,wangxy00}.
Similar $\gamma$-ray burst fireballs due to the color-flavor
locked phase-transition as well as magnetic field decay were noted
\citep{Ouyed2005,Ouyed2006}.
It is worth noting that \cite{Cui2007} showed statistically that
long-soft $\gamma$-ray burst could be really related to supernova
and that the asymmetry of bursts associated with supernova would
cause the kick of pulsars.

\acknowledgments

The authors would like to thank the referee for his/her very
constructive comments and suggestions, and wish to acknowledge
useful discussions with Guojun Qiao, Weiwei Zhu, Youling Yue and
Moda Cao at Peking University. We are indebted to J. Horvath for
many general discussions about supernova by forming bare quark
stars. This work is supported by NSFC (10573002, 10778611), the
Key Grant Project of Chinese Ministry of Education (305001), the
program of the Light in China's Western Region (LCWR, No.
LHXZ200602), and the Chun-Tsung Foundation.

\clearpage

\clearpage

\clearpage

\end{document}